\documentclass[aps, prb, twocolumn, showpacs, superscriptaddress]{revtex4-1}
\usepackage{subfigure}
\usepackage{amsmath,amssymb}
\usepackage{graphicx}
\usepackage{dcolumn}
\usepackage{bm}
\begin{document}

\title {Superfluid response in heavy fermion superconductors}
\author{Yin Zhong}
\email{zhongy05@hotmail.com}
\affiliation{Center for Interdisciplinary Studies $\&$ Key Laboratory for
Magnetism and Magnetic Materials of the MoE, Lanzhou University, Lanzhou 730000, China}
\author{Lan Zhang}
\affiliation{Center for Interdisciplinary Studies $\&$ Key Laboratory for
Magnetism and Magnetic Materials of the MoE, Lanzhou University, Lanzhou 730000, China}
\author{Can Shao}
\affiliation{Center for Interdisciplinary Studies $\&$ Key Laboratory for
Magnetism and Magnetic Materials of the MoE, Lanzhou University, Lanzhou 730000, China}
\author{Hong-Gang Luo}
\email{luohg@lzu.edu.cn}
\affiliation{Center for Interdisciplinary Studies $\&$ Key Laboratory for
Magnetism and Magnetic Materials of the MoE, Lanzhou University, Lanzhou 730000, China}
\affiliation{Beijing Computational Science Research Center, Beijing 100084, China}

\date{\today}
\begin{abstract}
Motivated by recent London penetration depth measurement [H. Kim et al. Phys. Rev. Lett. \textbf{114}, 027003 (2015)] and novel composite pairing scenario [O. Erten, R. Flint and P. Coleman, Phys. Rev. Lett. \textbf{114}, 027002 (2015)] on Yb-doped heavy fermion superconductor CeCoIn$_{5}$, we revisit the issue of superfluid response in microscopic heavy fermion lattice model. However, it is found that in literature explicit expression of superfluid response function in heavy fermion superconductor is rare. In this paper, we make a contribution to this issue by investigating superfluid density response function in celebrated Kondo-Heisenberg model. To be specific, we derive corresponding formalism from an effective fermionic large-N mean-field pairing Hamiltonian, whose pairing interaction is assumed to originate from effective local antiferromagnetic exchange interaction. Interestingly, it is found that physically correct superfluid density formula can only be obtained if external electromagnetic field is directly coupled to heavy fermion quasi-particle rather than the bare conduction electron or local moment. Such unique feature emphasizes the key role of Kondo-screening-renormalized heavy quasi-particle for low-temperature/energy thermodynamics and transport behaviors. As an important application, the theoretical result is compared to experimental measurement in heavy fermion superconductor CeCoIn$_{5}$ and Yb-doped Ce$_{1-x}$Yb$_{x}$CoIn$_{5}$, where the agreement is fairly good and the transition of pairing symmetry in the latter one is explained as a simple doping effect. In addition, the requisite formalism for the commonly encountered nonmagnetic impurity and non-local electrodynamic effect are developed. Inspired by the success in explaining classic $115$-series heavy
fermion superconductors, we expect the present theory can be applied to understand other heavy fermion superconductors like CeCu$_{2}$Si$_{2}$ and more generic multi-band superconductors.
\end{abstract}

\maketitle

\section{Introduction} \label{sec1}
Superfluid density or London penetration depth measurement is a fundamental experimental tool to detect low-temperature superconducting quasi-particle excitation in highly entangled unconventional pairing ground-state.\cite{Tinkham,Xiang,Poole} The low-temperature behavior of such quantity has provided invaluable information on the pairing symmetry for high-$T_{c}$ cuprate, iron pnictide and heavy fermion superconductors.\cite{Hardy1993,Kim2003,Prozorov,Ormeno2002,Ozcan2003,Chia2003,Hashimoto2013,Truncik2013,Shu2014,Kim2015}

In the field of heavy fermion compound, power-law temperature-dependence of superfluid density has been widely observed in the classic quasi-two-dimensional heavy fermion superconductor CeCoIn$_{5}$ with remarkably high $T_{c}=2.3\mathrm{K}$,\cite{Monthoux2001,Ormeno2002,Ozcan2003,Chia2003,Hashimoto2013,Truncik2013,Shu2014} although its detailed power-law behavior is still controversial due to proximity to possible magnetic quantum critical point.\cite{Ormeno2002,Ozcan2003,Chia2003,Hashimoto2013,Truncik2013}
When combining with other thermodynamic and transport measurements like heat capacity,\cite{Movshovich2001,An2010} thermal conductivity,\cite{Movshovich2001,Izawa2001} dc magnetization,\cite{Tayama2002} NMR,\cite{Kohori2001} tunneling spectrum,\cite{Ernst2010}
and the observation of a characteristic $(\pi,\pi,\pi)$ spin resonance mode and quasi-particle interference signal,\cite{Stock2008,Allan2013,Zhou2013,Morr2014} CeCoIn$_{5}$ is now believed to be a nodal-$d_{x^{2}-y^{2}}$ wave superconductor, which has remarkable similarity to the well-known high-$T_{c}$ cuprate.

More recently, London penetration depth measurement of Yb-doped CeCoIn$_{5}$,
i.e. Ce$_{1-x}$Yb$_{x}$CoIn$_{5}$, has been performed and unexpectedly the power-law behavior is gradually replaced by a fully gapped exponential behavior upon Yb-doping.\cite{Kim2015} (Note however a recent thermal conductivity measurement implies the nodal-d wave is robust and no pairing symmetry transition exists.\cite{Xu2015})
Such surprising experimental finding is explained to be a change of pairing symmetry
and even inspires an exotic local pairing scenario involving composite molecular superfluid and Lifshitz transition of nodal Fermi surface.\cite{Erten2015}

However, when we try to understand the superfluid response in those heavy fermion superconductors from a theoretical point, it is unfortunate to find that in literature explicit expression of superfluid response function, particularly for microscopic model Hamiltonian, is rare despite widespread and intensive experimental exploration. (Note however, some phenomenological Fermi liquid-based arguments and superfluid density formula of composite pairing indeed exist.\cite{Varma1986,Coleman1999}) So, in the present study, we focus on this issue by investigating superfluid density response function in the celebrated microscopic Kondo-Heisenberg model.\cite{Coleman1989} Specifically, we give a detailed derivation on corresponding formalism from an effective fermionic large-N mean-field pairing Hamitonian,\cite{Liu2012,Liu2014} whose pairing interaction is alternatively assumed to originate from effective local antiferromagnetic exchange interaction,\cite{Hu2012} though more conventional antiferromagnetic spin-fluctuation-exchange and resonance-valence-bond (RVB) theory give rise to identical effective pairing Hamiltonian.\cite{Coleman1989,Liu2012,Scalapino2012,Anderson2004,Lee2006,Zhong2015epjb}

\begin{figure}
\begin{center}
\includegraphics[width=0.6\columnwidth]{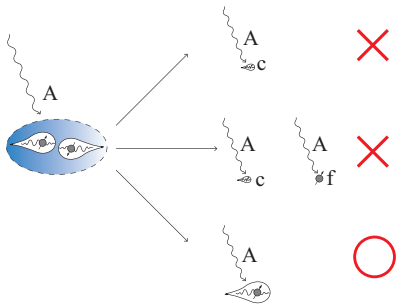}
\caption{\label{fig:ansatz} The external electromagnetic field $\vec{A}$ may couple to conduction electron ($c$-electron), auxiliary fermion ($f$-electron) and more directly the heavy fermion quasi-particle (large tadpole), only the last one is found to be responsible for realistic superfluid response in heavy fermion superconductors.}
\end{center}
\end{figure}

Importantly, we find only if external electromagnetic field is assumed to directly couple to heavy fermion quasi-particle rather than the bare conduction electron or local moment, the physically correct superfluid density formula can be obtained without any unsatisfactory breakdown. (See also Fig.~\ref{fig:ansatz}.) Generally, this feature underlies the fact that when considering low-temperature/energy thermodynamics and transport, the physical fermionic excitation should be the (composite) heavy fermion quasi-particle,\cite{Coleman} which results from nontrivial renormalization of collective Kondo screening effect. For superfluid density response considered in the present work, to match the temperature/doping evolution of superconducting heavy quasi-particle in superfluid density formula, characteristic quantities like effective velocity and mass have to be the ones of heavy quasi-particle, whose pairing drives the superconducting instability. These are the main findings of the present work and contribute to the active field of heavy fermion systems.

As a crucial application of the derived superfluid response formula, we revisit the undoped CeCoIn$_{5}$ and Yb-doped Ce$_{1-x}$Yb$_{x}$CoIn$_{5}$. For the former one, the agreement between theory and experiment is rather good, thus confirms the present theoretical formalism and underlies the dominating heavy fermion excitations. Furthermore, when applied to the Yb-doped latter one, we successfully explains the doping evolution of pairing symmetry in Ce$_{1-x}$Yb$_{x}$CoIn$_{5}$ as a simple doping effect by inspecting its pairing strength. The calculated temperature-dependent superfluid density in both nodal-d wave and nodeless-s wave states is well consistent with available measurement,\cite{Kim2015} which is a nontrivial check on our theory. In addition, we have derived related formulas for general nonmagnetic impurity and non-local electrodynamics effect, which may be important to realistic applications, e.g. La and Nd substituted CeCoIn$_{5}$.\cite{Kim2015}

Having succeed in explaining prototypical $115$-series heavy fermion superconductors,
we believe that the proposed formalism for superfluid response can be applied to understand other heavy fermion superconductors like the prototypical compound CeCu$_{2}$Si$_{2}$ and PrOs$_{4}$Sb$_{12}$ with possible time-reversal-symmetry breaking,\cite{Pfleiderer2009,Shu2009} if proper modification and extensions are included. More generally, it is expected our theory may also be useful for other multi-band superconductors, i.e. iron pnictide and LaOs$_{4}$Sb$_{12}$.\cite{Prozorov,Tee2012}

The remainder of this paper is organized as follows. In Sec. \ref{sec2}, Kondo-Heisenberg model is introduced and its mean-field normal-state Hamiltonian is derived. In Sec. \ref{sec3}, the pairing interaction is discussed and the corresponding pairing Hamiltonian is derived. In Sec. \ref{sec4}, three different superfluid density formulas are derived and analyzed. We focus on the application of our theory to heavy fermion superconductor CeCoIn$_{5}$ and its doped compound Ce$_{1-x}$Yb$_{x}$CoIn$_{5}$ in Sec.\ref{sec5}. In Sec. \ref{sec6}, both the impurity effect and non-local effect are discussed. Finally, Sec. \ref{sec7} is devoted to a brief conclusion.
\section{Kondo-Heisenberg model and normal state Hamiltonian}\label{sec2}
The definition of Kondo-Heisenberg model is standard, which reads,\cite{Coleman1989,Senthil2004}
\begin{eqnarray}
H=\sum_{k\sigma}\varepsilon_{k}c_{k\sigma}^{\dag}c_{k\sigma}+J_{K}\sum_{i}S_{i}^{c}\cdot S_{i}^{f}+J_{H}\sum_{<i,j>}S_{i}^{f}\cdot S_{j}^{f} \nonumber
\end{eqnarray}
For simplicity, the system is located on a regular square lattice and the resulting conduction electron band is $\varepsilon_{k}=-2t(\cos k_{x}+\cos k_{y})+4t'\cos k_{x}\cos k_{y}-\mu$ with chemical potential $\mu$. (Extension to other lattices like triangular or honeycomb lattice is straightforward.\cite{Zhong2012b,ZhangLan2015}) Next, we use fermionic representation to rewrite the degree of freedom of f-electron local spins as  $S_{i}^{f}=\frac{1}{2}\sum_{\sigma\sigma'}f_{i\sigma}^{\dag}\tau_{\sigma\sigma'}f_{i\sigma'}$
with $\tau$ being the standard Pauli matrices.
Moreover, to exclude nonphysical charge fluctuation of auxiliary fermion $f_{\sigma}$, it is crucial to enforce the local constraint $\sum_{\sigma}f_{i\sigma}^{\dag}f_{i\sigma}=1$ at each site.

Physically, this model describes two competing tendencies: The first is the Kondo screening, which leads to the formation of collective spin-singlet state among local moments and conduction electrons. The second is the short-ranged antiferromagnetic fluctuation introduced by Heisenberg interaction between local moments. It is believed that the observed complicated phenomena in diverse heavy electron systems can be captured by these two active factors.
Usually, to get qualitatively correct information in the paramagnetic heavy fermion liquid state, the fermionic large-N or slave-boson mean-field theory is
widely utilized.\cite{Read1983,Coleman1987} Here, we use the former one to get an effective mean-field Hamiltonian.

\subsection{Mean-field model in paramagnetic state}
After performing the standard large-N mean-field approximation,\cite{Lee2006,Senthil2004} we get the following Hamiltonian
\begin{eqnarray}
H&&=\sum_{k\sigma}[\varepsilon_{k}c_{k\sigma}^{\dag}c_{k\sigma}+\chi_{k}f_{k\sigma}^{\dag}f_{k\sigma}+\frac{J_{K}V}{2}(c_{k\sigma}^{\dag}f_{k\sigma}+f_{k\sigma}^{\dag}c_{k\sigma})]\nonumber\\
&&+E_{0}.\label{eq1}
\end{eqnarray}
Here, Kondo screening effect is encoded by an effective hybridization between conduction electron and local spins via $V=-\sum_{\sigma}\langle c_{i\sigma}^{\dag}f_{i\sigma}\rangle$. Meanwhile, local spins acquire dissipation $\chi_{k}=J_{H}\chi \eta_k + \lambda$ with $\eta_k = \cos k_{x}+\cos k_{y}$ due to the formation of nearest-neighbor valence-bond order $\chi=\sum_{\sigma}\langle f_{i\sigma}^{\dag}f_{j\sigma}\rangle$. Intuitively, such valence-bond order reflects the quantum dynamics of disordered local spins, which competes with magnetic long-ranged order. In addition, Lagrangian multiplier $\lambda$ is introduced to impose the local constraint on average and a constant energy shift $E_{0}=N_{s}[J_{K}V^{2}/2+J_{H}\chi^{2}-\lambda]$ with the number of lattice sites $N_{s}$ is added.

Using the following quasi-particle transformation relation
\begin{eqnarray}
&&c_{k\sigma}=\alpha_{k}A_{k\sigma}-\beta_{k}B_{k\sigma}\nonumber \\
&&f_{k\sigma}=\beta_{k}A_{k\sigma}+\alpha_{k}B_{k\sigma}\label{eq2}
\end{eqnarray}
with $\alpha_{k}^{2}=\frac{1}{2}(1+\frac{\varepsilon_{k}-\chi_{k}}{E_{0k}})$, $\beta_{k}^{2}=\frac{1}{2}(1-\frac{\varepsilon_{k}-\chi_{k}}{E_{0k}})$, $\alpha_{k}\beta_{k}=\frac{J_{K}V}{2E_{0k}}$ and $E_{0k}=\sqrt{(\varepsilon_{k}-\chi_{k})^{2}+(J_{K}V)^{2}}$, the original Hamiltonian Eq. (\ref{eq1}) is transformed into a diagonalized Hamiltonian
\begin{eqnarray}
H=\sum_{k\sigma}[E_{k}^{+}A_{k\sigma}^{\dag}A_{k\sigma}+E_{k}^{-}B_{k\sigma}^{\dag}B_{k\sigma}]+E_{0}\label{eq3}
\end{eqnarray}
with the quasi-particle energy $E_{k}^{\pm}=\frac{1}{2}(\varepsilon_{k}+\chi_{k}\pm E_{0k})$ for quasi-particle $A_{k\sigma}$ and $B_{k\sigma}$, respectively.

\section{Pairing interaction in Kondo-Heisenberg model}\label{sec3}
Because we are mainly interested in the superfluid response in corresponding superconducting pairing phase, a proper superconducting model is required to perform any realistic and sensible calculation.

For the present Kondo-Heisenberg model, it has long been proposed that the introduced Heisenberg interaction is able to mediate spin-singlet pairing between conduction electrons via the preformed pairing of auxiliary fermions.\cite{Coleman1989,Liu2012,Asadzadeh2014,Morr2014}
Microscopically, the underlying mechanism of preformed pairing of auxiliary fermions
can be attributed to the celebrated RVB state or alternatively the spin-fluctuation-exchange theory.\cite{Scalapino2012,Anderson2004,Lee2006,Coleman2010} However, to date, there is still no consensus on the true mechanism of such unconventional pairing and we are forced to reconsider this issue with more phenomenological approach.\cite{Coleman}

Theoretically, in both RVB or spin-fluctuation-exchange theory, the superconducting pairing is induced by certain effective antiferromagnetic interaction. The main difference is that the former emphasizes local real-space correlation while the latter one is focused on dynamic exchange of energy/momentum, i.e. the collective antiferromagnetic fluctuation.
In practice, since the observed pairing structure in heavy fermion superconductor has strong momentum-dependent behavior, one may expect that the driving force of pairing is the space-dependent part of effective pairing interaction rather than its time/energy part. In the other words, we may imagine a real-space pairing interaction and if we take the magnetic (fluctuation) nature of interaction into account, we will arrive at the following phenomenological local magnetic exchange interaction\cite{Hu2012}
\begin{eqnarray}
H_{int}=\sum_{ij}J_{ij}\vec{S}_{i}^{f}\cdot\vec{S}_{j}^{f}.\nonumber
\end{eqnarray}
Here, the spin density operator denotes the spin degree of freedom of local f-electron , which should result from the existing RKKY interaction. Furthermore, if only nearest-neighbor coupling is involved, this term naturally evolves into the Heisenberg term in Kondo-Heisenberg model. That is to say that the Kondo-Heisenberg model itself
has already included the basic element of superconducting pairing.

Now, we can write Heisenberg interaction term into its momentum-space version
as\cite{Coleman}
\begin{eqnarray}
H_{H}=\frac{1}{2}\sum_{ q}J_{q}\vec{S}_{q}^{f}\cdot \vec{S}_{-q}^{f}\nonumber
\end{eqnarray}
where $J_{q}=2J_{H}(\cos q_{x}+\cos q_{y})$ for square lattice.
Then, using the auxiliary fermionic representation of spin operator, the Heisenberg term is rewritten as
\begin{eqnarray}
H_{H}&&=\sum_{k,k',q}J_{q}(\frac{-1}{2}f_{k+q\uparrow}^{\dag}f_{k'-q\downarrow}^{\dag}f_{k'\downarrow}f_{k\uparrow}\nonumber\\
&&+\frac{\sigma\sigma'}{8}f_{k+q\sigma}^{\dag}f_{k'-q\sigma'}^{\dag}f_{k'\sigma'}f_{k\sigma}).\nonumber
\end{eqnarray}

Usually, it is well-known that in heavy fermion system the dominated magnetic correlation has antiferromagnetic feature, thus the spin-triplet pairing channel
is not favored and we will only consider the spin-singlet channel in our remaining parts. Therefore, the effective interaction in spin-singlet pairing channel is
\begin{eqnarray}
H_{H}=\sum_{k,k',q}\frac{-3}{4}J_{q}f_{k+q\uparrow}^{\dag}f_{k'-q\downarrow}^{\dag}f_{k'\downarrow}f_{k\uparrow}.\label{eq4}
\end{eqnarray}
On physical ground, such term describes that a pair of fermions with opposite spins feels a pairing interaction as $V_{q}=\frac{-3}{4}J_{q}$.

Then, let $q\rightarrow k-k'$ and symmetrize with momentum, we have the following pairing interaction\cite{Coleman}
\begin{eqnarray}
V_{k,k'}=\frac{1}{2}(V_{k-k'}+V_{k+k'})=-\frac{3}{2}J(\cos k_{x}\cos k_{x}'+\cos k_{y}\cos k_{y}'),\nonumber
\end{eqnarray}
which leads to an effective BCS pairing Hamiltonian as
\begin{eqnarray}
H_{pairing}=\sum_{k,k'}V_{k,k'}f_{k\uparrow}^{\dag}f_{-k\downarrow}^{\dag}f_{-k'\downarrow}f_{k'\uparrow}.\nonumber
\end{eqnarray}
Considering the C$_{4}$ rotation symmetry of square lattice,\cite{Kotliar1988} the present pairing interaction can be split into extended $s$-wave ($\gamma_{k}^{s}=\cos k_{x}+\cos k_{y}$) and $d_{x^{2}-y^{2}}$-wave ($\gamma_{k}^{d}=\cos k_{x}-\cos k_{y}$) pairing channel as
\begin{eqnarray}
&&V_{k,k'}=V_{k,k'}^{s}+V_{k,k'}^{d},\nonumber \\
&&V_{k,k'}^{s}=-\frac{3}{4}J_{H}\gamma_{k}^{s}\gamma_{k'}^{s},\nonumber \\
&&V_{k,k'}^{d}=-\frac{3}{4}J_{H}\gamma_{k}^{d}\gamma_{k'}^{d}. \nonumber
\end{eqnarray}
In other words, Heisenberg-like local antiferromagnetic exchange interaction may mediate extended-s and $d_{x^{2}-y^{2}}$-wave pairing on square lattice.

For each pairing symmetry, we obtain its pairing Hamiltonian as
\begin{eqnarray}
H_{pairing}=-\frac{3}{4}J_{H}\left(\sum_{k}\gamma_{k}f_{k\uparrow}^{\dag}f_{-k\downarrow}^{\dag}\right)\left(\sum_{k'}\gamma_{k'}f_{-k'\downarrow}f_{k'\uparrow}\right).\nonumber
\end{eqnarray}
Then, follow the spirit of classic BCS mean-field theory,\cite{Xiang} we introduce pairing order parameter
\begin{eqnarray}
\Delta=-\frac{3}{4}\sum_{k}\gamma_{k}\langle f_{k\uparrow}^{\dag}f_{-k\downarrow}^{\dag}\rangle=-\frac{3}{4}\sum_{k}\gamma_{k}\langle f_{-k\downarrow}f_{k\uparrow}\rangle,\nonumber
\end{eqnarray}
and the corresponding mean-field Hamiltonian reads
\begin{eqnarray}
H_{pairing}=J_{H}\Delta\sum_{k}\gamma_{k}(f_{k\uparrow}^{\dag}f_{-k\downarrow}^{\dag}
+f_{-k\downarrow}f_{k\uparrow})+\frac{4J_{H}\Delta^{2}}{3}.\label{eq5}
\end{eqnarray}
Note that such pairing Hamiltonian is identical to ones in
Refs.~\onlinecite{Liu2012,Liu2014,Asadzadeh2014}, where the RVB-like approach is utilized,\cite{Anderson2004} i.e. one assumes a spin-singlet pairing order of auxiliary fermion in real-space as
\begin{equation}
\Delta_{ij}=\langle f_{i\uparrow}^{\dag}f_{j\downarrow}^{\dag}-f_{i\downarrow}^{\dag}f_{j\uparrow}^{\dag}\rangle  \nonumber
\end{equation}

Obviously, although our method is more like the conventional antiferromagnetic spin-fluctuation-exchange theory, it is equivalent to the RVB-pairing theory at least in the present mean-field level, which emphasizes the core role of local magnetic interaction.\cite{Hu2012} In some sense, the mean-field pairing model misses the effect
of original pairing interaction, but if we include the fluctuation of pairing interaction, one may identify which microscopic interaction is responsible for pairing.

Now, adding this pairing Hamiltonian into the normal state model Eq.~\ref{eq1}, one can get the desirable model for heavy fermion superconductivity. In practice, the pairing between different heavy fermion quasi-particle bands is usually neglected due to the mismatch of Fermi surface as done in Ref.~\onlinecite{Morr2014}. Thus, the pairing Hamiltonian is simplified as
\begin{eqnarray}
H_{pairing}&&=J_{H}\Delta\sum_{k}\gamma_{k}(\beta_{k}^{2}(A_{k\uparrow}^{\dag}A_{-k\downarrow}^{\dag})
+\alpha_{k}^{2}(B_{k\uparrow}^{\dag}B_{-k\downarrow}^{\dag})\nonumber\\
&&+\alpha_{k}\beta_{k}(A_{k\uparrow}^{\dag}B_{-k\downarrow}^{\dag}+B_{k\uparrow}^{\dag}A_{-k\downarrow}^{\dag})+\mathrm{H}.\mathrm{c}.)+J_{H}\frac{4\Delta^{2}}{3}\nonumber\\
&&\simeq\sum_{k}[J_{H}\Delta_{k}^{A}(A_{k\uparrow}^{\dag}A_{-k\downarrow}^{\dag}+A_{-k\downarrow}A_{k\uparrow})\nonumber\\
&&+J_{H}\Delta_{k}^{B}(B_{k\uparrow}^{\dag}B_{-k\downarrow}^{\dag}+B_{-k\downarrow}B_{k\uparrow})]+J_{H}\frac{4\Delta^{2}}{3}.\label{eq6}
\end{eqnarray}

Here we have defined two gap functions $\Delta_{k}^{A}=\Delta\gamma_{k}\beta_{k}^{2}$ and $\Delta_{k}^{B}=\Delta\gamma_{k}\alpha_{k}^{2}$ for quasi-particle $A_{\sigma}$ and $B_{\sigma}$, respectively. Note that such gap function is modified by the normal-state coherent factor, whose momentum-dependence deviates from the pure pairing function $\gamma_{k}$ but the qualitative physics is unchanged by such modulation. Now, we see that the advantage of such simplification is to decouple each heavy fermion quasi-particle band, which leads to two independent BCS-like mean-field Hamiltonian. Therefore, after introducing Bogoliubov transformation for each quasi-particle band as
\begin{eqnarray}
&&A_{k\sigma}=\mu_{k}^{A}\tilde{A}_{k\sigma}-\nu_{k}^{A}\tilde{A}_{-k-\sigma}^{\dag}\nonumber\\
&&B_{k\sigma}=\mu_{k}^{A}\tilde{B}_{k\sigma}-\nu_{k}^{A}\tilde{B}_{-k-\sigma}^{\dag}\label{eq7}
\end{eqnarray}
with
\begin{eqnarray}
&&\mu_{k}^{A}=1-\nu_{k}^{A}=\frac{1}{2}(1+\frac{E_{k}^{+}}{E_{k}^{A}})\nonumber\\
&&\mu_{k}^{B}=1-\nu_{k}^{B}=\frac{1}{2}(1+\frac{E_{k}^{-}}{E_{k}^{B}}),\nonumber
\end{eqnarray}
the resultant Hamiltonian is
\begin{eqnarray}
H&&=\sum_{k}[E_{k}^{A}\tilde{A}_{k\sigma}^{\dag}\tilde{A}_{k\sigma}+E_{k}^{B}\tilde{B}_{k\sigma}^{\dag}\tilde{B}_{k\sigma}]\nonumber\\
&&+\sum_{k}(E_{k}^{+}+E_{k}^{-}-E_{k}^{A}-E_{k}^{B})+N_{s}E_{0}',\label{eq8}
\end{eqnarray}
where the energy spectrum of superconducting quasi-particle has the following form
\begin{eqnarray}
&&E_{k}^{A}=\sqrt{(E_{k}^{+})^{2}+(J_{H}\Delta_{k}^{A})^{2}}\nonumber\\
&&E_{k}^{B}=\sqrt{(E_{k}^{-})^{2}+(J_{H}\Delta_{k}^{B})^{2}}.\nonumber
\end{eqnarray}
The corresponding free energy and mean-field self-consistent equations are given in Appendix B.

\begin{figure}
\begin{center}
\includegraphics[width=0.6\columnwidth]{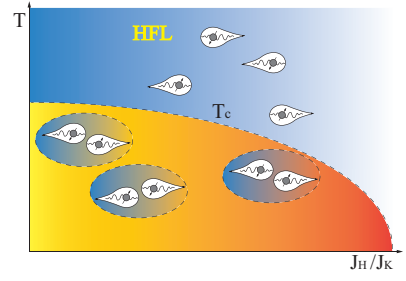}
\caption{\label{fig:SC} Above $T_{c}$, the itinerate heavy fermion quasi-particle forms heavy fermion liquid (HFL), which undergos a pairing instability into heavy fermion superconductor below $T_{c}$. When $J_{H}/J_{K}$ increases, the magnetic fluctuation is strengthened and magnetic or spin liquid states may ultimately dominate beyond critical ratio of $J_{H}/J_{K}$.}
\end{center}
\end{figure}
Now, the mean-field superconducting Hamiltonian for Kondo-Heisenberg model is obtained. When the temperature is above critical temperature $T_{c}$, there still exists the normal state described by Eq.~\ref{eq3}, where heavy fermion quasi-particle moves freely in the renormalized heavy Fermi liquid. If we cools the system below $T_{c}$, the normal heavy Fermi liquid transits into the heavy fermion superconducting state with pairing of heavy quasi-particle. To avoid the instability to quantum spin liquid or fractionalized Fermi liquid states,\cite{Senthil2004} the ratio of $J_{H}/J_{K}$ should not be larger than certain critical value. These are summarized in Fig.~\ref{fig:SC}

\section{Superfluid response formula}\label{sec4}
In this section, we will study the superfluid response function in heavy fermion superconductors. The main point of this section is for superfluid response in heavy fermion superconductors, which one is practically coupled to external electromagnetic field $\vec{A}$, conduction electron, auxiliary fermion or heavy fermion quasi-particle? As shown in the detailed discussion in this section, it is found that only the heavy fermion quasi-particle is most likely to couple with $\vec{A}$ strongly, which is also depicted in Fig.~\ref{fig:ansatz}.

\subsection{Superfluid response without auxiliary fermion}
For Kondo-Heisenberg model, by definition, the external electromagnetic potential $\vec{A}$ is exerted on conduction electron by substitution $t_{ij}\rightarrow t_{ij}e^{ieA_{ij}}$, where $t_{ij}$ is the hopping energy in real-space. Thus, the Hamiltonian is invariant under local U(1) charge gauge transformation via
\begin{eqnarray}
&&c_{\sigma}\rightarrow c_{\sigma}e^{i\theta_{i}},\nonumber\\
&&A_{ij}\rightarrow A_{ij}+(\theta_{i}-\theta_{j})/e\nonumber
\end{eqnarray}
while the auxiliary fermion
$f_{\sigma}$ does not directly response to $\vec{A}$ because it originates from the fermionic representation of spin and has no electric charge.\cite{Coleman}

Therefore, we assume that only conduction electron is involved in the calculation of superfluid response.

To discuss the superfluid response function, we consider conduction electron is coupled to electromagnetic field $\vec{A}_{q}$ via the following Hamiltonian\cite{Coleman}
\begin{eqnarray}
H'&&=\sum_{kq\sigma}e\vec{v}_{k}\cdot\vec{A}_{q}c_{k+q\sigma}^{\dag}c_{k\sigma}\nonumber\\
&&=\sum_{kq}e\vec{v}_{k}\cdot\vec{A}_{q}(c_{k+q\uparrow}^{\dag}c_{k\uparrow}+c_{k+q\downarrow}^{\dag}c_{k\downarrow})\nonumber\\
&&=\sum_{kq}e\vec{v}_{k}\cdot\vec{A}_{q}(c_{k+q\uparrow}^{\dag}c_{k\uparrow}+c_{-k\downarrow}c_{-k-q\downarrow}^{\dag})\nonumber\\
&&=\sum_{kq}e\vec{v}_{k}\cdot\vec{A}_{q}\psi_{k+q}^{\dag}\psi_{k}.\nonumber
\end{eqnarray}
Here, we have defined conduction electron velocity $\vec{v}_{k}=\frac{\partial \varepsilon_{k}}{\partial k_{i}}$ and the Nambu spinor $\psi_{k}^{\dag}=(c_{k\uparrow}^{\dag},c_{-k\downarrow})$. Note that the bare vertex is simply $ie\vec{v}_{k}$, which leads to paramagnetic current-current correlation function as
\begin{eqnarray}
\Pi(q)_{ij}^{p}&&=+e^{2}\sum_{k}v_{k}^{i}v_{k+q}^{j}Tr[G(k)G(k+q)]\nonumber\\
&&=+e^{2}\sum_{k}v_{k}^{i}v_{k+q}^{j}[G_{11}(k)G_{11}(k+q)+G_{12}(k)G_{21}(k+q)\nonumber\\
&&+(1\leftrightarrow2)],\nonumber
\end{eqnarray}
where $G$ is $2\times2$-matrix Nambu Green's function for conduction electron and its detail formalism is given in Appendix A.

Inserting the explicit expression of Green's function and setting zero-energy limit $i\Omega_{n}=0$ and zero-momentum limit $\vec{q}=0$, we get the contribution of paramagnetic current is
\begin{eqnarray}
Re\Pi(q=0)_{ij}^{p}=+2e^{2}\sum_{k}v_{k}^{i}v_{k}^{j}\left[\alpha_{k}^{4}\frac{\partial f_ {F}(E_{k}^{A})}{\partial E_{k}^{A}}+\beta_{k}^{4}\frac{\partial f_ {F}(E_{k}^{B})}{\partial E_{k}^{B}}\right].\nonumber
\end{eqnarray}

On the other hand, the diamagnetic current is coupled to $\vec{A}$ through
\begin{eqnarray}
H'&&=\sum_{kq\sigma}\frac{e^{2}}{2m}\vec{A}_{k'-q}^{\dag}\vec{A}_{k'}c_{k+q\sigma}^{\dag}c_{k\sigma}\nonumber\\
&&=\sum_{kq}\frac{e^{2}}{2m}\vec{A}_{k'-q}^{\dag}\cdot\vec{A}_{k'}(c_{k+q\uparrow}^{\dag}c_{k\uparrow}-c_{-k-q\downarrow}c_{-k\downarrow}^{\dag})\nonumber\\
&&=\sum_{kq}\frac{e^{2}}{2m}\vec{A}_{k'-q}^{\dag}\cdot\vec{A}_{k'}\psi_{k+q}^{\dag}\hat{\tau}_{z}\psi_{k}.\nonumber
\end{eqnarray}
The effective mass tensor for conduction electron is defined as $(\frac{1}{m})_{ij}=\frac{\partial^{2} \varepsilon_{k}}{\partial k_{i}\partial k_{j}}$. Here, the interacting vertex is $-(\frac{e^{2}}{2m})_{ij}\hat{\tau}_{z}$ and its corresponding zero-frequency diamagnetic current-current correlation is
\begin{eqnarray}
\Pi(q=0)_{ij}^{d}&&=+e^{2}\sum_{k}(\frac{1}{m})_{ij}Tr[\hat{\tau}_{z}G(k)]\nonumber\\
&&=e^{2}\sum_{k}(\frac{1}{m})_{ij}(G_{11}-G_{22}) \nonumber \\
&&=e^{2}\sum_{k}(\frac{-1}{m})_{ij}(\alpha_{k}^{2}\frac{E_{k}^{+}}{E_{k}^{A}}\tanh\frac{E_{k}^{A}}{2T}+\beta_{k}^{2}\frac{E_{k}^{-}}{E_{k}^{B}}\tanh\frac{E_{k}^{B}}{2T}). \nonumber
\end{eqnarray}

Combining the contribution from paramagnetic and diamagnetic part, we obtain
\begin{eqnarray}
Re\Pi(q=0)_{ij}&&=e^{2}\sum_{k}[\frac{-1}{m_{ij}}(\alpha_{k}^{2}\frac{E_{k}^{+}}{E_{k}^{A}}\tanh\frac{E_{k}^{A}}{2T}+\beta_{k}^{2}\frac{E_{k}^{-}}{E_{k}^{B}}\tanh\frac{E_{k}^{B}}{2T})\nonumber\\
&&+2v_{k}^{i}v_{k}^{j} (\alpha_{k}^{4}\frac{\partial f_ {F}(E_{k}^{A})}{\partial E_{k}^{A}}+\beta_{k}^{4}\frac{\partial f_ {F}(E_{k}^{B})}{\partial E_{k}^{B}})]. \nonumber
\end{eqnarray}

\begin{figure}
\begin{center}
\includegraphics[width=1.0\columnwidth]{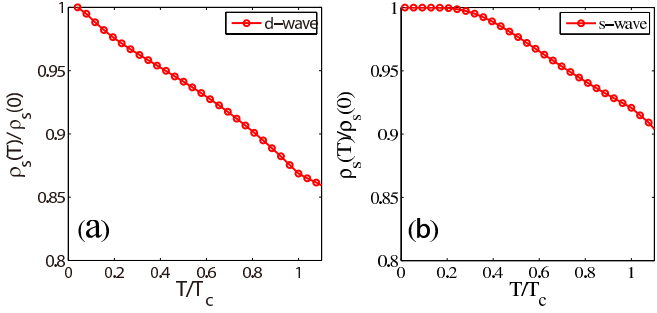}
\caption{\label{fig:SFf1} Normalized superfluid density $\rho_{s}(T)/\rho_{s}(0)$ versus normalized temperature T/T$_{c}$ for $d_{x^{2}-y^{2}}$ (a) and extended-s (b) wave pairing symmetry by using Eq.~\ref{eq9}.}
\end{center}
\end{figure}

Finally, by using of the definition of superfluid density $\frac{\rho_{ij}e^{2}}{m}=Re\Pi(q=0)_{ij}$, one gets the desirable superfluid density formula $\rho_{s}\equiv\rho_{ii}$,
\begin{eqnarray}
\frac{\rho_{s}(T)}{m}&&=\sum_{k}[\frac{\partial^{2}\varepsilon_{k}}{\partial k_{x}^{2}}(-\alpha_{k}^{2}\frac{E_{k}^{+}}{E_{k}^{A}}\tanh\frac{E_{k}^{A}}{2T}-\beta_{k}^{2}\frac{E_{k}^{-}}{E_{k}^{B}}\tanh\frac{E_{k}^{B}}{2T})\nonumber\\
&&+2(\frac{\partial\varepsilon_{k}}{\partial k_{x}})^{2}(\alpha_{k}^{4}\frac{\partial f_ {F}(E_{k}^{A})}{\partial E_{k}^{A}}+\beta_{k}^{4}\frac{\partial f_ {F}(E_{k}^{B})}{\partial E_{k}^{B}})]. \label{eq9}
\end{eqnarray}
Unfortunately, as what can be seen in Fig.~\ref{fig:SFf1}, for both $d_{x^{2}-y^{2}}$ and extended-s wave pairing symmetry, the superfluid density does not vanish at superconducting transition temperature T$_{c}$, thus the formula Eq.~\ref{eq9} is undoubtedly not valid. (The parameters we used are $t=1$, $t'=0.3$, $J_{K}=2$, $J_{H}=0.6$ and $n_{c}=0.9$, which are typical ones in the mean-field calculation of Kondo-Heisenberg model.\cite{Liu2012}) So, we have to answer why such formula is not correct although the derivation is rather straightforward and standard.

A careful reader may note that in the normal-state mean-field Hamiltonian Eq.~\ref{eq1}, conduction electron is hybridized with auxiliary fermion via the formation of Kondo screening ($V\neq0$ and is real), thus repeating the argument of U(1) charge gauge transformation discussed in the beginning of this subsection, the auxiliary fermion has to transform like
\begin{equation}
f_{i\sigma}\rightarrow f_{i\sigma}e^{i\theta_{i}},\nonumber
\end{equation}
which means that the auxiliary fermion $f_{\sigma}$ now effectively acquires electric charge due to the well-developed Kondo hybridization in heavy fermion liquid state.
As a matter of fact, this statement is more general and beyond the present mean-field level.\cite{Coleman} (In Appendix C, we give a brief discussion on this issue by path integral formalism.)

Therefore, when considering the superfluid response to external electromagnetic field, we should also include the contribution of auxiliary fermion and this will be analyzed  in next subsection.

\subsection{Superfluid response including auxiliary fermion}
In this section, we assume that beside the conduction electron, the auxiliary fermion also electromagnetically couples to $\vec{A}$, thus the total paramagnetic current is
\begin{equation}
j_{i}^{p}(q,\tau)=j_{i}^{c}(q,\tau)+j_{i}^{f}(q,\tau)\nonumber
\end{equation}
\begin{equation}
j_{i}^{c}(q,\tau)=e\sum_{k\sigma}v_{c}^{i}c_{k+q\sigma}^{\dag}(\tau)c_{k\sigma}(\tau)\nonumber
\end{equation}
\begin{equation}
j_{i}^{f}(q,\tau)=e\sum_{k\sigma}v_{f}^{i}f_{k+q\sigma}^{\dag}(\tau)f_{k\sigma}(\tau).\nonumber
\end{equation}
Here, $v_{c}^{i}=\partial\varepsilon_{k}/\partial k_{i}$ and  $v_{f}^{i}=\partial\chi_{k}/\partial k_{i}$ are the effective velocity of conduction electron and auxiliary fermion, respectively.

Therefore, the paramagnetic response function is
\begin{eqnarray}
\Pi_{ij}(q,\tau)&&=-\langle T_{\tau}j_{i}^{p}(q,\tau)j_{i}^{p}(-q,0)\rangle\nonumber\\
&&=\Pi_{ij}^{cc}(q,\tau)+\Pi_{ij}^{cf}(q,\tau)+\Pi_{ij}^{fc}(q,\tau)+\Pi_{ij}^{ff}(q,\tau)\nonumber.
\end{eqnarray}
Note that there exist four terms, among which $\Pi_{ij}^{cc}$ and $\Pi_{ij}^{ff}$ denote the contribution of conduction electron and auxiliary fermion, respectively. The left terms $\Pi_{ij}^{cc}$ and $\Pi_{ij}^{ff}$ are the cross-term, which implies the quantum interference effect between different currents. In last subsection, we have calculated $\Pi_{ij}^{cc}$,
\begin{eqnarray}
Re\Pi(q=0)_{ii}^{cc}=2e^{2}\sum_{k}(v_{c}^{i})^{2}[\alpha_{k}^{4}\frac{\partial f_ {F}(E_{k}^{A})}{\partial E_{k}^{A}}+\beta_{k}^{4}\frac{\partial f_ {F}(E_{k}^{B})}{\partial E_{k}^{B}}].\nonumber
\end{eqnarray}

After a long but straightforward calculation, we obtain
\begin{eqnarray}
Re\Pi(q=0)_{ii}^{ff}=2e^{2}\sum_{k}(v_{f}^{i})^{2}[\beta_{k}^{4}\frac{\partial f_ {F}(E_{k}^{A})}{\partial E_{k}^{A}}+\alpha_{k}^{4}\frac{\partial f_ {F}(E_{k}^{B})}{\partial E_{k}^{B}}].\nonumber
\end{eqnarray}

Since $Re\Pi(q=0)_{ii}^{cf}=Re\Pi(q=0)_{ii}^{fc}$, the remaining work is to calculate $\Pi(q=0)_{ii}^{cf}$, whose expression is
\begin{eqnarray}
\Pi(q)_{ii}^{cf}&&=e^{2}\sum_{k}v_{c}^{i}v_{f}^{i}
[G_{13}(k)G_{31}(k+q)+G_{14}(k)G_{41}(k+q)\nonumber\\
&&+(1\leftrightarrow2)].\nonumber
\end{eqnarray}
After inserting the explicit form of $G_{13}$ and $G_{14}$, the final result reads
\begin{eqnarray}
Re\Pi(q=0)_{ii}^{cf}=2e^{2}\sum_{k}v_{c}^{i}v_{f}^{i}[\alpha_{k}^{2}\beta_{k}^{2}\frac{\partial f_ {F}(E_{k}^{A})}{\partial E_{k}^{A}}+\alpha_{k}^{2}\beta_{k}^{2}\frac{\partial f_ {F}(E_{k}^{B})}{\partial E_{k}^{B}}].\nonumber
\end{eqnarray}

So, the paramagnetic current response is found to be
\begin{eqnarray}
Re\Pi(q=0)_{ii}^{p}&&=2e^{2}\sum_{k}[(\alpha_{k}^{4}(v_{c}^{i})^{2}+\beta_{k}^{4}(v_{f}^{i})^{2}+2\alpha_{k}^{2}\beta_{k}^{2}v_{c}v_{f})\nonumber\\
&&\times\frac{\partial f_ {F}(E_{k}^{A})}{\partial E_{k}^{A}}
+(\beta_{k}^{4}(v_{c}^{i})^{2}+\alpha_{k}^{4}(v_{f}^{i})^{2}+2\alpha_{k}^{2}\beta_{k}^{2}v_{c}v_{f})\nonumber\\
&&\times\frac{\partial f_ {F}(E_{k}^{B})}{\partial E_{k}^{B}}]\nonumber
\end{eqnarray}
and the diamagnetic current response is easy to obtain as
\begin{eqnarray}
Re\Pi(q=0)_{ii}^{d}&&=-e^{2}\sum_{k}
[\alpha_{k}^{2}(\frac{1}{m_{c}})_{ii}+\beta_{k}^{2}(\frac{1}{m_{f}})_{ii} ]\frac{E_{k}^{+}}{E_{k}^{A}}\tanh\frac{E_{k}^{A}}{2T}\nonumber\\
&&+[\beta_{k}^{2}(\frac{1}{m_{c}})_{ii}+\alpha_{k}^{2}(\frac{1}{m_{f}})_{ii} ]\frac{E_{k}^{-}}{E_{k}^{B}}\tanh\frac{E_{k}^{B}}{2T}. \nonumber
\end{eqnarray}

Therefore, our wanted superfluid density formula is
\begin{equation}
\frac{\rho_{s}(T)}{m}=\frac{1}{e^{2}}(Re\Pi(q=0)_{ii}^{d}+Re\Pi(q=0)_{ii}^{p}).\label{eq10}
\end{equation}

\begin{figure}
\begin{center}
\includegraphics[width=1.0\columnwidth]{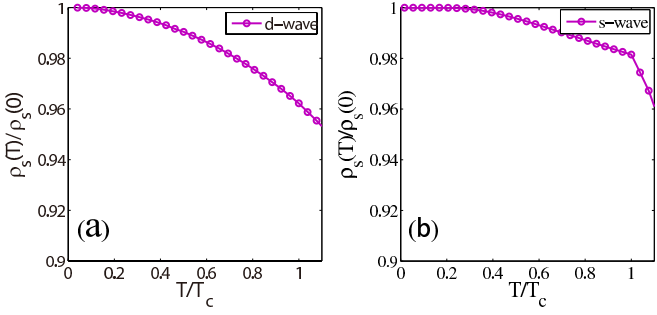}
\caption{\label{fig:SFf2} Normalized superfluid density $\rho_{s}(T)/\rho_{s}(0)$ versus normalized temperature T/T$_{c}$ for $d_{x^{2}-y^{2}}$ (a) and extended-s (b) wave pairing symmetry by using Eq.~\ref{eq10}.}
\end{center}
\end{figure}

In Fig.~\ref{fig:SFf2}, we show the calculation data of Eq.~\ref{eq10}, which again exhibits incorrect result. (The parameters are the same to the ones in last subsection.) The reason for this may be due to the mismatch of temperature-dependence between bare particles ($v_{c}$, $v_{f}$, $m_{c}$, $m_{f}$) and the renormalized heavy fermion quasi-particle ($E_{k}^{A}$, $E_{k}^{B}$). In other words, the bare and renormalized particles have rather different temperature dependence, thus the paramagnetic part cannot cancel out the diamagnetic term when $T>T_{c}$.
More generally, when calculating physical quantities at $T=0$, since no temperature effect is involved, the underlying breakdown of the corresponding formalism is hidden but at finite temperature, one may encounter the mentioned problem.

After all, we indeed confront a serious problem on the calculation of superfluid density.

\subsection{A useful and correct superfluid response formula}
\begin{figure}
\begin{center}
\includegraphics[width=1.0\columnwidth]{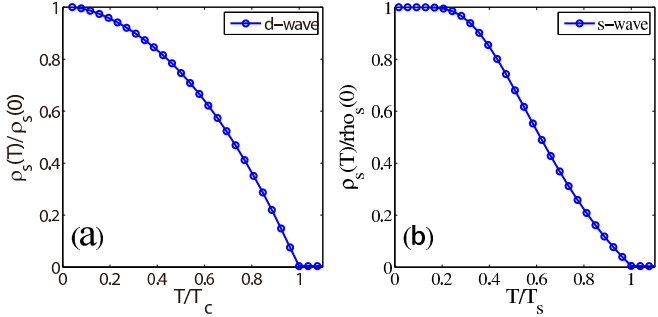}
\caption{\label{fig:SFf3} Normalized superfluid density $\rho_{s}(T)/\rho_{s}(0)$ versus normalized temperature T/T$_{c}$ for $d_{x^{2}-y^{2}}$ (a) and extended-s (b) wave pairing symmetry by using Eq.~\ref{eq11}.}
\end{center}
\end{figure}
In last section, we have seen that the standard process of calculating superfluid response in fact breaks down. However, there exists a hint to resolve this question by noting that when collective Kondo screening develops, the quasi-particle of the system is the heavy fermion quasi-particle $A_{k\sigma}$ and $B_{k\sigma}$. If all interesting physics occurs at energy-scale smaller than the width of quasi-particle band $E_{k}^{+}$ and $E_{k}^{-}$, the heavy fermion quasi-particle can be seen as the only fundamental or true particles in our problem, thus we may expect the low-energy thermodynamics and transport are dominated by those heavy quasi-particle alone. For heavy fermion superconductivity, it is commonly believed that the pairing results from heavy quasi-particle but not directly involving with original conduction electron.\cite{Poole,Monthoux2001,Coleman,Pfleiderer2009} Therefore, by considering these two factors, we propose that the heavy fermion quasi-particle is directly responsible for superfluid electromagnetic response, thus the corresponding paramagnetic current is
\begin{equation}
j_{i}^{p}(q,\tau)=e\sum_{k\sigma}(v_{A}^{i}A_{k+q\sigma}^{\dag}(\tau)A_{k\sigma}(\tau)+v_{B}^{i}B_{k+q\sigma}^{\dag}(\tau)B_{k\sigma}(\tau))\nonumber
\end{equation}
where the effective velocity for heavy fermion quasi-particle $A_{\sigma}$ and $B_{\sigma}$ are defined by $v_{A}^{i}=\frac{\partial E_{k}^{+}}{\partial k_{i}}$ and $v_{B}^{i}=\frac{\partial E_{k}^{-}}{\partial k_{i}}$, respectively. At the same time, the diamagnetic current is found to be
\begin{eqnarray}
j_{i}^{d}(q,\tau)&&=e^{2}A_{j}\sum_{k\sigma}((\frac{1}{m_{A}})_{ij}A_{k+q\sigma}^{\dag}(\tau)A_{k\sigma}(\tau)\nonumber\\
&&+(\frac{1}{m_{B}})_{ij}B_{k+q\sigma}^{\dag}(\tau)B_{k\sigma}(\tau)).\nonumber
\end{eqnarray}

Then, with the same treatment in last subsection, one can use these two heavy quasi-particle currents to derive Eq.~\ref{eq11}.
\begin{eqnarray}
\frac{\rho_{s}(T)}{m}&&=\sum_{k}[-\frac{\partial^{2}E_{k}^{+}}{\partial k_{x}^{2}}\frac{E_{k}^{+}}{E_{k}^{A}}\tanh\frac{E_{k}^{A}}{2T}+2(\frac{\partial E_{k}^{+}}{\partial k_{x}})^{2}\frac{\partial f_ {F}(E_{k}^{A})}{\partial E_{k}^{A}}\nonumber\\
&&-\frac{\partial^{2}E_{k}^{-}}{\partial k_{x}^{2}}\frac{E_{k}^{-}}{E_{k}^{B}}\tanh\frac{E_{k}^{B}}{2T}
+2(\frac{\partial E_{k}^{-}}{\partial k_{x}})^{2}\frac{\partial f_ {F}(E_{k}^{B})}{\partial E_{k}^{B}}], \label{eq11}
\end{eqnarray}
which is a useful and physically correct superfluid density formula and is the most important finding of the present paper.
As shown in Fig.~\ref{fig:SFf3}, the superfluid density is correctly vanished when transition temperature $T_{c}$ is approached from heavy fermion superconducting phase.
Meanwhile, the global behavior of temperature-dependence in $\rho_{s}(T)$ is also consistent with general physical expectation, e.g. extended-s wave should gives rise to a flat $T$-dependence while the nodal $d_{x^{2}-y^{2}}$-wave pairing usually leads to a power-law behavior in low-temperature regime.\cite{Xiang,Poole}

\begin{figure}
\begin{center}
\includegraphics[width=0.5\columnwidth]{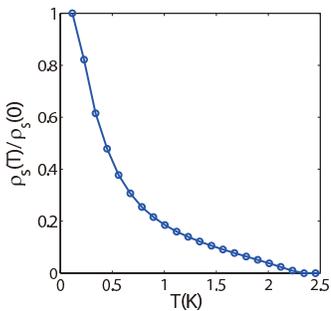}
\caption{\label{fig:CeCoIn5} Normalized superfluid density $\rho_{s}(T)/\rho_{s}(0)$ versus temperature T for heavy fermion superconductor CeCoIn$_{5}$ by using Eq.~\ref{eq11}.}
\end{center}
\end{figure}

We should emphasize that our formula Eq.~\ref{eq11} is not only valid in superconducting mean-field model Eq.\ref{eq8} but can be used in other heavy fermion superconducting models. As an example, we consider the effective model proposed in Ref.\onlinecite{Morr2014}, which is acquired from fitting to quasi-particle interference experiments in CeCoIn$_{5}$.\cite{Allan2013,Zhou2013} The details of their model is not given in this paper and we refer the reader to their original article, here only the calculation result is shown in Fig.~\ref{fig:CeCoIn5}.

When comparing to the experimental data of undoped CeCoIn$_{5}$ in Ref.~\onlinecite{Ormeno2002}, (See also Fig.~\ref{fig:CeCoIn5}.) we see that the calculated $\rho_{s}(T)$ does not agree with the global behavior in the experimental measurement. This may be due to the fact that the model parameters in Ref.~\onlinecite{Morr2014} are not temperature-dependent, thus certain physical quantities like superfluid density cannot be faithfully calculated. Therefore, it is important to include the thermal effect and we will revisit this issue in next section.

\section{Superfluid response in heavy fermion superconductor CeCoIn$_{5}$ and Ce$_{1-x}$Yb$_{x}$CoIn$_{5}$}\label{sec5}

\subsection{Superfluid response in heavy fermion superconductor CeCoIn$_{5}$}

\begin{figure}
\includegraphics[width=1.0\columnwidth]{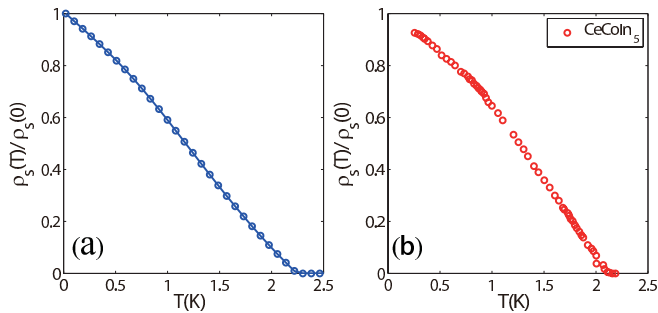}
\caption{\label{fig:10} (a) Calculated normalized superfluid density $\rho_{s}(T)/\rho_{s}(0)$ in the superconducting state versus $T$.
(b) The normalized superfluid density $\rho_{s}(T)/\rho_{s}(0)$ of CeCoIn$_{5}$ in Ref.~\onlinecite{Ormeno2002}.}
\end{figure}

\begin{figure}
\begin{center}
\includegraphics[width=0.8\columnwidth]{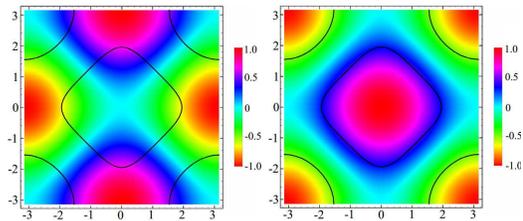}
\caption{\label{fig:pairingstrength} Mean-field Fermi surface (bold black line), $d_{x^{2}-y^{2}}$ (Left) and extended-s (Right) wave pairing function in square lattice Brillouin zone.}
\end{center}
\end{figure}

In this section, we proceed to study the superfluid response in a kind of typical heavy fermion superconductor CeCoIn$_{5}$. To meet with the realistic experimental measurement, we fix model parameters as $t=-1$, $t'=0.3$, $J_{K}=2$, $J_{H}=0.6$, $V=-0.2965$, $\chi=0.2222$ and $n_{c}=0.9$,\cite{Zhong2015epjb} thus the resulting normal state Fermi surface is shown in Fig.~\ref{fig:pairingstrength}, which mimics the multi-Fermi surface structure observed in ARPES measurement.\cite{Koitzsch2009,Jia2011}

Performing calculation with Eq.~\ref{eq11} and assuming $d_{x^{2}-y^{2}}$-wave pairing, we compare our theoretical result to experimental data in Fig.~\ref{fig:10}. From Fig.~\ref{fig:10} we see that our theory is good consistent with the global behavior in microwave surface impedance measurements of Ref.~\onlinecite{Ormeno2002} and also agrees with our previous work.\cite{Zhong2015epjb} Particularly, the observed low-temperature linear/power-law behavior of superfluid density is well reproduced by our theoretical model with only assumption of $d_{x^{2}-y^{2}}$-wave symmetry.\cite{Ormeno2002,Ozcan2003,Chia2003,Hashimoto2013,Truncik2013}

\subsection{Issues on Ce$_{1-x}$Yb$_{x}$CoIn$_{5}$}
\begin{figure}
\begin{center}
\includegraphics[width=0.9\columnwidth]{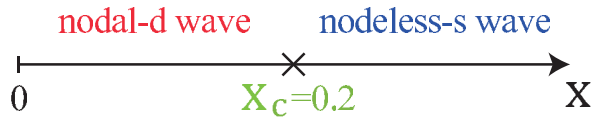}
\caption{\label{fig:x1} Change of pairing symmetry from nodal-d to nodeless-s wave at critical Yb-doping $x_{c}=0.2$ for Ce$_{1-x}$Co$_{x}$In$_{5}$.}
\end{center}
\end{figure}

Next, it is interesting to understand the possible pairing symmetry transition in Yb-doped CeCoIn$_{5}$, which is suggested by recent superfluid density measurement and inspires an exotic local electron pairing scenario.\cite{Kim2015,Erten2015}
Effectively, Yb-doping adds extra electron carrier into CeCoIn$_{5}$, thus density of charge carrier is modified as $n_{c}(x)=0.9+0.8x$ with $x$ being nominal doping.\cite{Erten2015} It is found in Ref.~\onlinecite{Kim2015} that when $x>0.2$, the low-temperature London penetration depth shows a full gap behavior ($\Delta \lambda(T)\sim T^{n}$ with
$n>2$). Therefore, such anomalous behavior of superfluid response strongly suggests that the superconducing pairing symmetry should change from nodal to nodeless structure. (See also Fig.~\ref{fig:x1}.)

Motivated by above observation, authors in Ref.~\onlinecite{Erten2015} propose that
the change of pairing symmetry may result from the formation of composite pairing, i.e. electrons in two orthogonal orbits screening the same local f-electron spin to form pairing. In this scenario, Yb-doping leads to a Lifshitz transition of nodal Fermi surface into a fully gapped composite-pairing molecular superfluid, thus the nodeless behavior observed in superfluid density measurement is expected.

Here we would like to propose an alternative possibility, i.e. the change of pairing structure may be just a doping effect but not related to emergence of exotic electron pairing. To be specific, we utilize our normal-state model in Sec.~\ref{sec2} and calculate the doping-dependence of the so-called pairing strength invented in Ref.~\onlinecite{Hu2012},
\begin{eqnarray}
\lambda^{\ast}\equiv\sum_{k\in FS}\gamma_{k}^{2}\nonumber.
\end{eqnarray}
Here, the dimensionless $\lambda^{\ast}$ measures the overlap of pairing function $\gamma_{k}$ on normal-state Fermi surface. Obviously, less (more) overlap between nodal point/line (maximized gap zone) and Fermi surface is helpful to lower ground-state energy and the corresponding pairing strength is larger. In other words, pairing symmetry with larger $\lambda^{\ast}$ will ultimately become the true superconducting pairing instability.
For the Fermi surface shown in Fig.~\ref{fig:pairingstrength}, although the nodal line of $d_{x^{2}-y^{2}}$-wave has much overlap over Fermi surface, its high gap zone also has large overlap, thus the whole effect is to promote
$d_{x^{2}-y^{2}}$-wave over than the nodeless extended-s wave, (The nodal line of extended-s wave indeed exists but it does not contact with the present Fermi surface, thus it behaves like a nodeless pairing.) which is consistent with the calculated value $\lambda_{d}^{\ast}=0.957>\lambda_{s}^{\ast}=0.875$ and as well agrees with the observed dominated nodal d-wave feature in CeCoIn$_{5}$.

\begin{figure}
\begin{center}
\includegraphics[width=1.0\columnwidth]{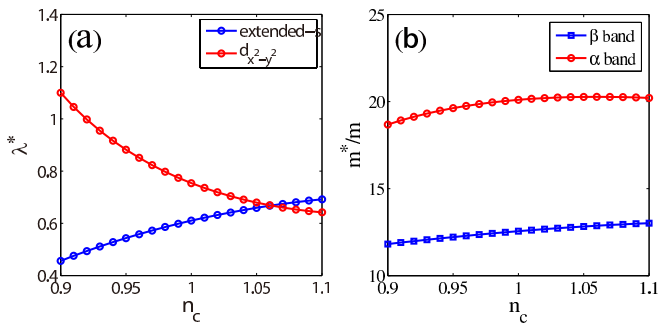}
\caption{\label{fig:20} Doping evolution of pairing strength $\lambda^{\ast}$ (a) and effective mass ratio $m^{\ast}/m$ (b).}
\end{center}
\end{figure}

When considering the doping case, we plot the pairing strength $\lambda^{\ast}$ and
effective mass ratio $m^{\ast}/m$ in Fig.~\ref{fig:20}. (We use $t=-1$, $t'=0.14t$, $J_{K}=1.92$, $J_{H}=0.6$, $\chi=0.2222$ and $T=0.01$. $V$, $\mu$ and $\lambda$ are self-consistently solved through mean-field equations.)
It is found that when $n_{c}<1.06$, ($n_{c}=1.06$ corresponds to the observed critical doping $x_{c}=0.2$.) $d_{x^{2}-y^{2}}$-wave pairing dominates over
extended-s wave, but if $n_{c}>1.06$, the situation is inverted and the effectively nodeless extended-s wave is the leading pairing symmetry. Therefore, this simple calculation indeed indicates there may be a transition at $n_{c}=1.06$ ($x_{c}=0.2$) from nodal d-wave to nodeless s-wave symmetry. When checking the doping evolution of effective mass, it can be seen that there is no clear signal of any singular behavior in both $\alpha$ ($E_{k}^{+}$) and $\beta$ ($E_{k}^{-}$) heavy quasi-particle bands, thus we expect the changes of electronic band or Lifshitz transition observed in ARPES and dHvA measurements may not be an active factor for the pairing symmetry transition, but it is an issue by itself and needs further studies.\cite{Dudy2013,Polyakov2012}

\begin{figure}
\begin{center}
\includegraphics[width=1.0\columnwidth]{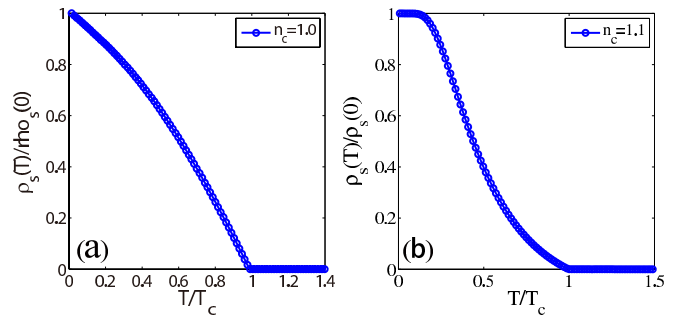}
\caption{\label{fig:30}  The superfluid density of nodal-d wave (a) and nodeless-s wave (b) for $n_{c}=1.0$ ($x=0.125<x_{c}=0.2$) and $n_{c}=1.10$ ($x=0.25>x_{c}=0.2$), respectively.}
\end{center}
\end{figure}

Furthermore, Fig.~\ref{fig:30} shows the temperature-dependent superfluid density in both nodal d-wave and nodeless s-wave state for $n_{c}=1.0$ ($x=0.125<x_{c}=0.2$) and $n_{c}=1.10$ ($x=0.25>x_{c}=0.2$), respectively.
Importantly, the behaviors of calculated results are consistent with the experimental data in Ref.~\onlinecite{Kim2015}, where low-temperature behavior of London penetration depth is changed from power-law to exponential when doping is over $x_{c}=0.2$. In
addition, it is noted that the pairing symmetry transition here is likely a first-order transition since the corresponding derivative of ground-state energy is discontinuous at the putative transition point $x=0.2$. Thus, no radical change of observable like $T_{c}$ is expected around this transition, which is also consistent with the doping evolution of Yb in Ref.~\onlinecite{Kim2015}.

Before ending this section, it is interesting to note that for the heavy fermion superconductor CeRhIn$_{5}$,\cite{Hegger2000} where coexistence of antiferromagnetism and superconductivity is firmly established,\cite{Park2006} the present superfluid response formula may also be applicable if we use a four-quasi-particle energy-band model resulted from magnetic and pairing orders. Thus, we should sum the contributions from all of four bands with corresponding quasi-particle effective mass and velocity.
It will be a good test for our theory to fit to the future London penetration depth measurement in this compound.

\section{Nonmagnetic impurity effect and non-local effect on superfluid response }\label{sec6}
\subsection{Nonmagnetic impurity effect}
For realistic materials, the nonmagnetic impurity scattering is unavoidable and is ubiquitous. Particularly, it plays a key role in the low temperature regime where linear-$T$ dependence of nodal d-wave is replaced by the $T^{2}$ law.\cite{Goldenfeld1993} Here, we give the corresponding formalism, which may be relevant to explaining realistic experimental data.

The main point of nonmagnetic impurity scattering is that it introduces effective damping rate $\Gamma_{A}$ and $\Gamma_{B}$ for each heavy quasi-particle. Then, the corresponding retarded heavy quasi-particle Green's function in superconducting state reads
\begin{eqnarray}
G_{AA}^{R}(k,\omega)=\left(
         \begin{array}{cc}
           G_{11}^{AA} & G_{12}^{AA}\\
           G_{21}^{AA} & G_{22}^{AA} \\
         \end{array}
       \right),\nonumber
\end{eqnarray}
where
\begin{eqnarray}
&&G_{11}^{AA}=\frac{(\mu_{k}^{A})^{2}}{\omega-E_{k}^{A}+i\Gamma_{A}}+\frac{(\nu_{k}^{A})^{2}}{\omega+E_{k}^{A}+i\Gamma_{A}} \nonumber \\
&&G_{12}^{AA}=\mu_{k}^{A}\nu_{k}^{A}(\frac{1}{\omega-E_{k}^{A}+i\Gamma_{A}}-\frac{1}{\omega+E_{k}^{A}+i\Gamma_{A}}) \nonumber\\
&&G_{21}^{AA}=\mu_{k}^{A}\nu_{k}^{A}(\frac{1}{\omega-E_{k}^{A}+i\Gamma_{A}}-\frac{1}{\omega+E_{k}^{A}+i\Gamma_{A}}) \nonumber\\
&&G_{22}^{AA}=\frac{(\nu_{k}^{A})^{2}}{\omega-E_{k}^{A}+i\Gamma_{A}}+\frac{(\mu_{k}^{A})^{2}}{\omega+E_{k}^{A}+i\Gamma_{A}}\nonumber
\end{eqnarray}
and the case for $G_{BB}^{R}(k,\omega)$ is similar.
Next, follow the same treatment in the main text, we obtain the expression of superfluid density as
\begin{eqnarray}
\frac{\rho_{s}(T)}{m}&&=\sum_{k}[\frac{\partial^{2}E_{k}^{+}}{\partial k_{x}^{2}}\frac{E_{k}^{+}}{E_{k}^{A}}W_{k}^{A}+2(\frac{\partial E_{k}^{+}}{\partial k_{x}})^{2}Z_{k}^{A}\nonumber\\
&&+\frac{\partial^{2}E_{k}^{-}}{\partial k_{x}^{2}}\frac{E_{k}^{-}}{E_{k}^{B}}W_{k}^{B}
+2(\frac{\partial E_{k}^{-}}{\partial k_{x}})^{2}Z_{k}^{B}]. \nonumber
\end{eqnarray}
with
\begin{eqnarray}
&&W_{k}^{A}=\int_{-\infty}^{\infty}d\omega f_{F}(\omega)(\frac{\Gamma_{A}/\pi}{(\omega-E_{k}^{A})^{2}+\Gamma_{A}^{2}}-\frac{\Gamma_{A}/\pi}{(\omega+E_{k}^{A})^{2}+\Gamma_{A}^{2}})
\nonumber\\
&&W_{k}^{B}=\int_{-\infty}^{\infty}d\omega f_{F}(\omega)(\frac{\Gamma_{B}/\pi}{(\omega-E_{k}^{B})^{2}+\Gamma_{B}^{2}}-\frac{\Gamma_{B}/\pi}{(\omega+E_{k}^{B})^{2}+\Gamma_{B}^{2}})\nonumber\\
&&Z_{k}^{A}=-\frac{1}{\pi}\int_{-\infty}^{\infty}d\omega f_{F}(\omega)Im \frac{(\omega+i\Gamma_{A})^{2}+(E_{k}^{A})^{2}}{[(\omega+i\Gamma_{A})^{2}-(E_{k}^{A})^{2}]^{2}}
\nonumber\\
&&Z_{k}^{B}=-\frac{1}{\pi}\int_{-\infty}^{\infty}d\omega f_{F}(\omega)Im \frac{(\omega+i\Gamma_{B})^{2}+(E_{k}^{B})^{2}}{[(\omega+i\Gamma_{B})^{2}-(E_{k}^{B})^{2}]^{2}}
.\nonumber
\end{eqnarray}

By using the above equations, one can inspect the explicit behavior of superfluid density in all temperature regime. Here, we investigate the low temperature behavior by expanding $Z_{k}^{\alpha=A,B}$ as\cite{Xiang} ($W_{k}^{\alpha}$ is approximated as a constant in this case)
\begin{equation}
Z_{k}^{\alpha}=-\frac{1}{\pi}[\frac{\Gamma_{\alpha}}{\Gamma_{\alpha}^{2}+(E_{k}^{\alpha})^{2}}-\frac{\Gamma_{\alpha}(\Gamma_{\alpha}^{2}-3(E_{k}^{\alpha})^{2})T^{2}}{3(\Gamma_{\alpha}^{2}+(E_{k}^{\alpha})^{2})^{3}}+O(T^{4})]\nonumber
\end{equation}
where the first term basically denotes a correction to zero-temperature superfluid density and the second term introduces the impurity-dominated $T^{2}$ dependence since this term vanishes with $\Gamma_{\alpha}\rightarrow0$.

Then, for nodal superconductors like $d_{x^{2}-y^{2}}$-wave in undoped CeCoIn$_{5}$,
it is found that the low-temperature superfluid density behaves as
\begin{eqnarray}
\rho_{s}(T)&&\simeq\rho_{A}(0)(1-\frac{2\Gamma_{A}}{\pi\Delta_{A}}\ln\frac{2\Delta_{A}}{\Gamma_{A}}-\frac{T^{2}}{3\pi\Gamma_{A}\Delta_{A}})
\nonumber\\
&&+\rho_{B}(0)(1-\frac{2\Gamma_{B}}{\pi\Delta_{B}}\ln\frac{2\Delta_{B}}{\Gamma_{B}}-\frac{T^{2}}{3\pi\Gamma_{B}\Delta_{B}})\nonumber
\end{eqnarray}
which is in contrast to the linear-$T$ behavior of the pure sample. Here, $\rho_{\alpha}(0)$ and $\Delta_{\alpha}$ are the zero-temperature superfluid density and effective gap for each quasi-particle band. We should emphasize that both the pure nodal-d wave and the one with impurity are at least qualitative consistent with the observed power-law behaviors in CeCoIn$_{5}$. For La and Nd substituted CeCoIn$_{5}$,\cite{Kim2015} this impurity induced $T^{2}$ dirty d-wave behavior has also been observed by using a tunnel-diode resonator technique.
Furthermore, for the possible nodeless pairing state in Yb-doped Ce$_{1-x}$Yb$_{x}$CoIn$_{5}$, no qualitative behaviors are changed by nonmagnetic impurity scattering since all low-energy excitations are fully gapped.

\subsection{Non-local effect}
For nodal pairing states like the studied $d_{x^{2}-y^{2}}$-wave, when approaching gap nodal points, its coherent length $\xi\sim v_{F}/\Delta$ is larger than its London penetration depth $\lambda$, thus the so-called non-local effects are important,\cite{Kosztin1997} which may be responsible for the observed power-law behavior of London penetration depth in CeCoIn$_{5}$.\cite{Ormeno2002,Chia2003,Hashimoto2013}

Follow Ref.\onlinecite{Kosztin1997}, we find that for a specular boundary, the temperature-dependent London
penetration depth reads
\begin{eqnarray}
\lambda(T)=\frac{2}{\pi}\int_{0}^{\infty}\frac{dq}{q^{2}+Re\Pi(q,\omega=0)}\nonumber
\end{eqnarray}
where the momentum-dependent superfluid response function
\begin{eqnarray}
Re\Pi(q,\omega=0)&&=-e^{2}\sum_{k}[\frac{\partial^{2}E_{k}^{+}}{\partial k_{x}^{2}}\frac{E_{k}^{+}}{E_{k}^{A}}\tanh\frac{E_{k}^{A}}{2T}\nonumber\\
&&+(\mu_{k}^{A}\mu_{k+q}^{A}+\nu_{k}^{A}\nu_{k+q}^{A})^{2}
\frac{\tanh\frac{\beta E_{k}^{A}}{2}-\tanh\frac{\beta E_{k+q}^{A}}{2}}{E_{k}^{A}-E_{k+q}^{A}}\nonumber\\
&&+(\mu_{k}^{A}\nu_{k+q}^{A}-\nu_{k}^{A}\mu_{k+q}^{A})^{2}
\frac{\tanh\frac{\beta E_{k}^{A}}{2}+\tanh\frac{\beta E_{k+q}^{A}}{2}}{E_{k}^{A}+E_{k+q}^{A}}\nonumber\\
&&+\frac{\partial^{2}E_{k}^{-}}{\partial k_{x}^{2}}\frac{E_{k}^{-}}{E_{k}^{B}}\tanh\frac{E_{k}^{B}}{2T}\nonumber\\
&&+(\mu_{k}^{B}\mu_{k+q}^{B}+\nu_{k}^{B}\nu_{k+q}^{B})^{2}
\frac{\tanh\frac{\beta E_{k}^{B}}{2}-\tanh\frac{\beta E_{k+q}^{B}}{2}}{E_{k}^{B}-E_{k+q}^{B}}\nonumber\\
&&+(\mu_{k}^{B}\nu_{k+q}^{B}-\nu_{k}^{B}\mu_{k+q}^{B})^{2}
\frac{\tanh\frac{\beta E_{k}^{B}}{2}+\tanh\frac{\beta E_{k+q}^{B}}{2}}{E_{k}^{B}+E_{k+q}^{B}}]\nonumber.
\end{eqnarray}

At low temperature ($T<T^{\ast}\sim\xi/\lambda(0)\Delta$), for nodal superconducting states, the above equation gives $\triangle\lambda(T)\equiv\lambda(T)-\lambda(0)\sim T^{2}$, which is similar to the effect of nonmagnetic impurity and modifies the linear-$T$ behavior. We note authors in Ref.\onlinecite{Chia2003} argued that the penetration depth of CeCoIn$_{5}$ is governed by this nonlocal electrodynamics effect
but rather than the nonmagnetic impurity effect.
\section{conclusion and discussion} \label{sec7}
In this work, we have systematically studied superfluid density response formula in pure sample and with nonmagnetic impurity. These formulas have been successfully applied to undoped heavy fermion superconductor CeCoIn$_{5}$ and its Yb-doped counterpart Ce$_{1-x}$Yb$_{x}$CoIn$_{5}$. The agreement between theory and experiment in those typical heavy fermion superconductors is rather good, which confirms our basic formalism and emphasizes the core role of heavy fermion quasi-particle. The present work may provide useful formalism to understand and explain existing and future superfluid density experiments in heavy fermion superconductivity.

In the main text, we have noted that the measurement of thermal conductivity suggests there is no pairing symmetry transition in Ce$_{1-x}$Yb$_{x}$CoIn$_{5}$, which is in contrast to the result of London penetration depth experiment. Frankly, it is still unable to judge which experiment gives true physical behavior, but if we use the model in Ref.~\onlinecite{Morr2014} with our superfluid density formula, we find only nodal-d wave is possible in this case.

It is intriguing to note the possibility that if the critical quasi-particle picture is still valid in heavy fermion superconducting states with proximity to quantum critical point,\cite{Abrahams2014} the paramagnetic current should contribute an extra $T^{1/2}$ due to the critical enhancement of effective velocity. Then, based on our superfluid response formula, the nodal-d wave states may have the $T^{3/2}$ behavior, which agrees well with existing data.\cite{Ormeno2002,Chia2003,Hashimoto2013}

In addition, as suggested in Ref.\onlinecite{Coleman1999,Erten2015}, since both intersite and onsite two-channel effects are probably important for heavy fermion superconductivity, the composite pairing has to be considered seriously, which is an interesting and crucial issue for future study.

\begin{acknowledgments}
Y. Z. thank stimulating discussion with Jianhui Dai and Piers Coleman, who brings us the issue of non-local effect and composite pairing in heavy fermion superconductors. The work was supported partly by NSFC, PCSIRT (Grant No. IRT1251), the Fundamental Research Funds for the Central Universities and the national program for basic research of China.
\end{acknowledgments}

\appendix

\section{Single particle Green's function}
In the superconducting state, the single particle imaginary time Green's function is defined as follows£º
\begin{eqnarray}
G(k,\tau)&&\equiv-\langle T_{\tau}(c_{k\uparrow}(\tau),c_{-k\downarrow}^{\dag}(\tau),f_{k\uparrow}(\tau),f_{-k\downarrow}^{\dag}(\tau))\otimes\left(
                           \begin{array}{c}
                             c_{k\uparrow}^{\dag} \\
                             c_{-k\downarrow}   \\
                             f_{k\uparrow}^{\dag} \\
                             f_{-k\downarrow} \\
                           \end{array}
                         \right)\rangle \nonumber \\
&&= \left(
      \begin{array}{cccc}
        G_{11} & G_{12} & G_{13} & G_{14} \\
        G_{21} & G_{22} & G_{23} & G_{24} \\
        G_{31} & G_{32} & G_{33} & G_{34} \\
        G_{41} & G_{42} & G_{43} & G_{44} \\
      \end{array}
    \right)
.\nonumber
\end{eqnarray}

With the superconducting mean-field Hamiltonian, the matrix element in momentum-energy representation reads
\begin{eqnarray}
G_{11}(k)&&=\alpha_{k}^{2}[\frac{(\mu_{k}^{A})^{2}}{i\omega_{n}-E_{k}^{A}}+\frac{(\nu_{k}^{A})^{2}}{i\omega_{n}+E_{k}^{A}}] \nonumber \\
&&+\beta_{k}^{2}[\frac{(\mu_{k}^{B})^{2}}{i\omega_{n}-E_{k}^{B}}+\frac{(\nu_{k}^{B})^{2}}{i\omega_{n}+E_{k}^{B}}]\nonumber\\
\nonumber
\end{eqnarray}

\begin{eqnarray}
G_{12}(k)=G_{21}(k)&&=\alpha_{k}^{2}\mu_{k}^{A}\nu_{k}^{A}[\frac{1}{i\omega_{n}-E_{k}^{A}}-\frac{1}{i\omega_{n}+E_{k}^{A}}]\nonumber \\
&&+\beta_{k}^{2}\mu_{k}^{B}\nu_{k}^{B}[\frac{1}{i\omega_{n}-E_{k}^{B}}-\frac{1}{i\omega_{n}+E_{k}^{B}}]\nonumber
\end{eqnarray}

\begin{eqnarray}
G_{22}(k)&&=\alpha_{k}^{2}[\frac{(\nu_{k}^{A})^{2}}{i\omega_{n}-E_{k}^{A}}+\frac{(\mu_{k}^{A})^{2}}{i\omega_{n}+E_{k}^{A}}]\nonumber \\
&&+\beta_{k}^{2}[\frac{(\nu_{k}^{B})^{2}}{i\omega_{n}-E_{k}^{B}}+\frac{(\mu_{k}^{B})^{2}}{i\omega_{n}+E_{k}^{B}}]\nonumber
\end{eqnarray}

\begin{eqnarray}
G_{13}(k)=G_{31}(k)&&=\alpha_{k}\beta_{k}[\frac{(\mu_{k}^{A})^{2}}{i\omega_{n}-E_{k}^{A}}+\frac{(\nu_{k}^{A})^{2}}{i\omega_{n}+E_{k}^{A}}]\nonumber\\
&&-\alpha_{k}\beta_{k}[\frac{(\mu_{k}^{B})^{2}}{i\omega_{n}-E_{k}^{B}}+\frac{(\nu_{k}^{B})^{2}}{i\omega_{n}+E_{k}^{B}}]\nonumber
\end{eqnarray}

\begin{eqnarray}
G_{14}(k)=G_{41}(k)&&=G_{23}(k)=G_{32}(k)\nonumber\\
&&=\alpha_{k}\beta_{k}\mu_{k}^{A}\nu_{k}^{A}[\frac{1}{i\omega_{n}-E_{k}^{A}}-\frac{1}{i\omega_{n}+E_{k}^{A}}]\nonumber\\
&&-\alpha_{k}\beta_{k}\mu_{k}^{B}\nu_{k}^{B}[\frac{1}{i\omega_{n}-E_{k}^{B}}-\frac{1}{i\omega_{n}+E_{k}^{B}}]\nonumber
\end{eqnarray}

\begin{eqnarray}
G_{24}(k)=G_{42}(k)&&=\alpha_{k}\beta_{k}[\frac{(\nu_{k}^{A})^{2}}{i\omega_{n}-E_{k}^{A}}+\frac{(\mu_{k}^{A})^{2}}{i\omega_{n}+E_{k}^{A}}]\nonumber\\
&&-\alpha_{k}\beta_{k}[\frac{(\nu_{k}^{B})^{2}}{i\omega_{n}-E_{k}^{B}}+\frac{(\mu_{k}^{B})^{2}}{i\omega_{n}+E_{k}^{B}}].\nonumber
\end{eqnarray}

\begin{eqnarray}
G_{33}(k)&&=\beta_{k}^{2}[\frac{(\mu_{k}^{A})^{2}}{i\omega_{n}-E_{k}^{A}}+\frac{(\nu_{k}^{A})^{2}}{i\omega_{n}+E_{k}^{A}}]\nonumber\\
&&+\alpha_{k}^{2}[\frac{(\mu_{k}^{B})^{2}}{i\omega_{n}-E_{k}^{B}}+\frac{(\nu_{k}^{B})^{2}}{i\omega_{n}+E_{k}^{B}}]\nonumber
\end{eqnarray}

\begin{eqnarray}
G_{34}(k)=G_{43}(k)&&=\beta_{k}^{2}\mu_{k}^{A}\nu_{k}^{A}[\frac{1}{i\omega_{n}-E_{k}^{A}}-\frac{1}{i\omega_{n}+E_{k}^{A}}]\nonumber\\
&&+\alpha_{k}^{2}\mu_{k}^{B}\nu_{k}^{B}[\frac{1}{i\omega_{n}-E_{k}^{B}}-\frac{1}{i\omega_{n}+E_{k}^{B}}]\nonumber
\end{eqnarray}

\begin{eqnarray}
G_{44}(k)&&=\beta_{k}^{2}[\frac{(\nu_{k}^{A})^{2}}{i\omega_{n}-E_{k}^{A}}+\frac{(\mu_{k}^{A})^{2}}{i\omega_{n}+E_{k}^{A}}]\nonumber\\
&&+\alpha_{k}^{2}[\frac{(\nu_{k}^{B})^{2}}{i\omega_{n}-E_{k}^{B}}+\frac{(\mu_{k}^{B})^{2}}{i\omega_{n}+E_{k}^{B}}].\nonumber
\end{eqnarray}

\section{Free energy and mean-field equation in heavy fermion superconducting state}
Using the diagnolized mean-field Hamiltonian Eq.\ref{eq8}, one can readily get the free energy as
\begin{eqnarray}
F&&=-2T\sum_{k}[\ln(1+e^{-\beta E_{k}^{A}})+\ln(1+e^{-\beta E_{k}^{B}})]\nonumber\\
&&+\sum_{k}(\varepsilon_{k}+\chi_{k}-E_{k}^{A}-E_{k}^{B})+N_{s}E_{0}'\nonumber
\end{eqnarray}

Then, minimizing the free energy function versus mean-field parameters $V$, $\chi$, $\lambda$, $\mu$ and $\Delta$, we obtain the following five self-consistent equations:
\begin{eqnarray}
2=J_{K}\sum_{k}\frac{1}{E_{0k}}(\frac{E_{k}^{+}}{E_{k}^{A}}\tanh\frac{\beta E_{k}^{A}}{2}-\frac{E_{k}^{-}}{E_{k}^{B}}\tanh\frac{\beta E_{k}^{B}}{2})\nonumber
\end{eqnarray}
\begin{eqnarray}
2\chi=J_{K}\sum_{k}\eta_{k}[\beta_{k}^{2}\frac{E_{k}^{+}}{E_{k}^{A}}\tanh\frac{\beta E_{k}^{A}}{2}+\alpha_{k}^{2}\frac{E_{k}^{-}}{E_{k}^{B}}\tanh\frac{\beta E_{k}^{B}}{2}]\nonumber
\end{eqnarray}
\begin{eqnarray}
0=\sum_{k}[2\beta_{k}^{2}\frac{E_{k}^{+}}{E_{k}^{A}}\tanh\frac{\beta E_{k}^{A}}{2}+2\alpha_{k}^{2}\frac{E_{k}^{-}}{E_{k}^{B}}\tanh\frac{\beta E_{k}^{B}}{2}]\nonumber
\end{eqnarray}
\begin{eqnarray}
2(n_{c}-1)=-\sum_{k}[2\alpha_{k}^{2}\frac{E_{k}^{+}}{E_{k}^{A}}\tanh\frac{\beta E_{k}^{A}}{2}+2\beta_{k}^{2}\frac{E_{k}^{-}}{E_{k}^{B}}\tanh\frac{\beta E_{k}^{B}}{2}]\nonumber
\end{eqnarray}
\begin{eqnarray}
\Delta^{2}=J_{H}\sum_{k}[\frac{(\Delta_{k}^{A})^{2}}{2E_{k}^{A}}\tanh\frac{\beta E_{k}^{A}}{2}+\frac{(\Delta_{k}^{B})^{2}}{2E_{k}^{B}}\tanh\frac{\beta E_{k}^{B}}{2}].\nonumber
\end{eqnarray}

\section{Charged auxiliary fermion}
In this section, we follow the argument of Ref.\onlinecite{Coleman} to explain the fact that auxiliary fermion has acquired electric charge via Kondo screening in heavy fermion liquid state.

In literature, the finite temperature or imaginary-time path integral formalism is often used and the corresponding formula is\cite{Coleman1989,Senthil2004,Coleman}
\begin{eqnarray}
Z=\int \mathcal{D}\bar{V}\mathcal{D}VD\lambda \mathcal{D}\bar{\chi}\mathcal{D}\chi\mathcal{D}c^{\dag}\mathcal{D} c \mathcal{D}f^{\dag}\mathcal{D} f e^{-S},\nonumber
\end{eqnarray}
and the action reads
\begin{eqnarray}
S&&=\int_{0}^{\beta}\{\sum_{k\alpha}c_{k\alpha}^{\dag}(\partial_{\tau}+\varepsilon_{k})c_{k\alpha} +\sum_{i\alpha}f_{i\alpha}^{\dag}(\partial_{\tau}+i\lambda_{i})f_{i\alpha}\nonumber\\
&&+\frac{J_{K}}{2}\sum_{i\alpha}[\bar{V}_{i}(c_{i\alpha}^{\dag}f_{i\alpha})+V_{i}(f_{i\alpha}^{\dag}c_{i\alpha})]\nonumber\\
&&+\frac{J_{H}}{2}\sum_{\langle ij\rangle\alpha}[\bar{\chi}_{ij}(f_{i\alpha}^{\dag}f_{j\alpha})+\chi_{ij}(f_{j\alpha}^{\dag}f_{i\alpha})]\nonumber\\
&&+N(J_{K}\sum_{i}\frac{|V_{i}|^{2}}{4}+J_{H}\sum_{\langle ij\rangle}\frac{|\chi_{ij}|^{2}}{4}-i\sum_{i}\lambda_{i} q)\}.\nonumber
\end{eqnarray}
Here $q=1/N$ and $N=2$ for the physical spin case and all mean-field order parameters are now dynamic fields. Evidently, the dynamic Lagrangian parameter $\lambda$ enforces the constraint if we integrate it out. Meanwhile, integrating out valence-bond order $\chi_{ij}$ will recover the original Heisenberg interaction term. Thus, the above path integral formalism is an exact description of original Kondo-Heisenberg model.\cite{Coleman,Senthil2004}

Then, set $V_{i}=V_{i}e^{i\phi_{i}}$ with the new $V_{i}$ being a real field and absorb the phase part into $f_{i}$, this gives a substitute $\lambda_{i}\rightarrow\lambda_{i}+\partial_{\tau}\phi_{i}$ and the Kondo hybridizing field $V_{i}$ is real.
\begin{eqnarray}
S&&=\int_{0}^{\beta}\{\sum_{k\alpha}c_{k\alpha}^{\dag}(\partial_{\tau}+\varepsilon_{k})c_{k\alpha} +\sum_{i\alpha}f_{i\alpha}^{\dag}(\partial_{\tau}+i(\lambda_{i}+\partial_{\tau}\phi_{i})f_{i\alpha}\nonumber\\
&&+\frac{J_{K}}{2}\sum_{i\alpha}V_{i}[c_{i\alpha}^{\dag}f_{i\alpha}+f_{i\alpha}^{\dag}c_{i\alpha}]\nonumber\\
&&+\frac{J_{H}}{2}\sum_{\langle ij\rangle\alpha}[\bar{\chi}_{ij}(f_{i\alpha}^{\dag}f_{j\alpha})+\chi_{ij}(f_{j\alpha}^{\dag}f_{i\alpha})]\nonumber\\
&&+N(J_{K}\sum_{i}\frac{V_{i}^{2}}{4}+J_{H}\sum_{\langle ij\rangle}\frac{|\chi_{ij}|^{2}}{4}-i\sum_{i}\lambda_{i} q)\}.\nonumber
\end{eqnarray}

As emphasized in Ref.\onlinecite{Coleman}, when Kondo screening develops, the internal gauge symmetry is lost in the above action due to the Anderson-Higgs mechanism, which states that the the absorb of gapless phase into gauge field leads to the mass of gauge field itself. In other words, $f$-fermion appears a physical object and is free of internal gauge structure.

Furthermore, we can define a new $\lambda_{i}$ to absorb $\partial_{\tau}\phi_{i}$,
\begin{eqnarray}
S&&=\int_{0}^{\beta}\{\sum_{k\alpha}c_{k\alpha}^{\dag}(\partial_{\tau}+\varepsilon_{k})c_{k\alpha} +\sum_{i\alpha}f_{i\alpha}^{\dag}(\partial_{\tau}+i\lambda_{i})f_{i\alpha}\nonumber\\
&&+\frac{J_{K}}{2}\sum_{i\alpha}V_{i}[c_{i\alpha}^{\dag}f_{i\alpha}+f_{i\alpha}^{\dag}c_{i\alpha}]\nonumber\\
&&+\frac{J_{H}}{2}\sum_{\langle ij\rangle\alpha}[\bar{\chi}_{ij}(f_{i\alpha}^{\dag}f_{j\alpha})+\chi_{ij}(f_{j\alpha}^{\dag}f_{i\alpha})]\nonumber\\
&&+N(J_{K}\sum_{i}\frac{V_{i}^{2}}{4}+J_{H}\sum_{\langle ij\rangle}\frac{|\chi_{ij}|^{2}}{4}-i\sum_{i}\lambda_{i}q)\}.\nonumber
\end{eqnarray}

Here, we should note the integral measurement of $\lambda_{i}$ is changed from $\{-\infty,\infty\}$ to complex plane, thus the number of degree of freedom is unchanged.
\begin{eqnarray}
\int_{-\infty}^{\infty} \frac{d\lambda_{i}}{2\pi}[...]\rightarrow \int_{c} \frac{d\lambda_{i}}{2\pi}[...]\nonumber.
\end{eqnarray}

Now, under the electrodynamic gauge transformation, $c_{i\alpha}\rightarrow c_{i\alpha}e^{i\theta_{i}}$, we have to use $f_{i\alpha}\rightarrow f_{i\alpha}e^{i\theta_{i}}$ since $V_{i}$ is real and charge-less. This clearly endorse the physical charge into composite $f$-fermion and it can response to the external electromagnetic field as true physical electron excitation.\cite{Coleman} After all, the full action including external electromagnetic field is given by
\begin{eqnarray}
S&&=\int_{0}^{\beta}\{\sum_{ij\alpha}c_{i\alpha}^{\dag}(\partial_{\tau}\delta_{ij}+t_{ij}e^{ieA_{ij}})c_{j\alpha} +\sum_{i\alpha}f_{i\alpha}^{\dag}(\partial_{\tau}+i\lambda_{i})f_{i\alpha}\nonumber\\
&&+\frac{J_{K}}{2}\sum_{i\alpha}V_{i}[c_{i\alpha}^{\dag}f_{i\alpha}+f_{i\alpha}^{\dag}c_{i\alpha}]\nonumber\\
&&+\frac{J_{H}}{2}\sum_{\langle ij\rangle\alpha}[\bar{\chi}_{ij}e^{-ieA_{ij}}(f_{i\alpha}^{\dag}f_{j\alpha})+\chi_{ij}e^{ieA_{ij}}(f_{j\alpha}^{\dag}f_{i\alpha})]\nonumber\\
&&+N(J_{K}\sum_{i}\frac{V_{i}^{2}}{4}+J_{H}\sum_{\langle ij\rangle}\frac{|\chi_{ij}|^{2}}{4}-i\sum_{i}\lambda_{i}q)\},\nonumber
\end{eqnarray}
where both conduction electron and auxiliary fermion are coupled to external electromagnetic field.

\end{document}